\documentclass[11pt]{article}
\usepackage{a4wide}                      
\usepackage{epsf}                        
\usepackage{latexsym}                    
\usepackage{amsfonts}                    
\usepackage{amssymb}                     

\newcommand{\sect}[1]{\setcounter{equation}{0}\section{#1}}
\renewcommand{\theequation}{\arabic{section}.\arabic{equation}}

\def\be{\begin{equation}}
\def\ee{\end{equation}}
\def\bea{\begin{eqnarray}}
\def\eea{\end{eqnarray}}

\def\nnw{\nonumber \\ [.2cm]}

\def\hsp#1{\hspace*{#1}}

\def\part{\partial}

\def\ux{{\underline x}}
\def\hD{\hat D}
\def\hV{\hat V}
 
\def\hchi{{\hat{\chi}}}

\def\hmu{{\hat{\mu}}}                  
\def\hnu{{\hat{\nu}}}                  
\def\hrho{{\hat{\rho}}}                
  
\def\ha{{\hat a}}     
\def\hb{{\hat b}}     
\def\hzeta{{\hat{\zeta}}}    

\def\mn{{\mu\nu}}

\def\hmn{{\hmu\hnu}}

\def\makeatletter{\catcode`\@=11}
\makeatletter
\def\mathbox#1{\hbox{$\m@th#1$}}%
\def\math@ccstyles#1#2#3#4#5#6#7{{\leavevmode
      \setbox0\mathbox{#6#7}%
      \setbox2\mathbox{#4#5}%
      \dimen@ #3%
      \baselineskip\z@\lineskiplimit#1\lineskip\z@
      \vbox{\ialign{##\crcr
             \hfil \kern #2\box2 \hfil\crcr
             \noalign{\kern\dimen@}%
             \hfil\box0\hfil\crcr}}}}
\def\mathaccstyles{\math@ccstyles\maxdimen}
\def\maththroughstyles{\math@ccstyles{-\maxdimen}}
\def\unity%
 {\maththroughstyles{.45\ht0}\z@\displaystyle {\mathchar"006C}\displaystyle 1}
\begin{document}

\rightline{UG-FT-267/10}
\rightline{CAFPE-137/10}
\rightline{May 2010}
\vspace{1.5truecm}

\centerline{\huge \bf The group structure of non-Abelian}
\vspace{.6truecm}
\centerline{\huge \bf NS-NS transformations}
\vspace{1.3truecm}

\centerline{
    {\large \bf Bert Janssen}\footnote{E-mail address: 
                                  {\tt bjanssen@ugr.es} }
    {\bf and} 
    {\large \bf Airam Marcos-Caballero}\footnote{E-mail address: 
                                  {\tt airam.mrc@gmail.com}}
                                                            }
\vspace{.4cm}
\centerline{{\it Departamento de F\'{\i}sica Te\'orica y del Cosmos and}}
\centerline{{\it Centro Andaluz de F\'{\i}sica de Part\'{\i}culas Elementales}}
\centerline{{\it Universidad de Granada, 18071 Granada, Spain}}

\vspace{2truecm}

\centerline{\bf ABSTRACT}
\vspace{.5truecm}

\noindent
We study the transformations of the worldvolume fields of a system of multiple coinciding D-branes
under gauge transformations of the supergravity Kalb-Ramond field. We find that the pure 
gauge part of these NS-NS transformations can be written as a $U(N)$ symmetry of the underlying
Yang-Mills group, but that in general the full NS-NS variations get mixed up non-trivially with 
the $U(N)$. We compute the commutation relations and the Jacobi identities of the bigger group 
formed by the NS-NS and $U(N)$ transformations.

\newpage
\sect{Introduction}

Although the new physics associated with systems of multiple coinciding D-branes has 
been known for more than a decade and a half \cite{Witten}, there are still a few open issues 
concerning the 
effective action that describes the low-energy dynamics of the system. One of the famous unsolved 
problems is the construction of a non-Abelian Born-Infeld action (see for example \cite{Tseytlin2}
and references thereof). The form of the Chern-Simons action is much better known, through the
work of \cite{Douglas}-\cite{Myers}. Still, also here are a few open issues left, such as the 
gauge invariance of the action.  The invariance under $U(N)$ gauge symmetry is straightforward to
show, as the entire action is written in terms of objects that transform in the adjoint 
representation and is traced over all Yang-Mills indices. However, the Chern-Simons term is also 
the one responsible for the couplings of the multiple D-brane system to the background fields of 
supergravity and the invariance of this part of the action under the gauge transformations of the 
supergravity background field has drawn little or no attention.

A few early attempts to prove the gauge invariance were made in \cite{GHT, Ciocarlie}, but the 
first syste\-ma\-tic approach was done is a series of papers by J. Adam et al., proving the 
explicit invariance under Ramond-Ramond (R-R) and massive gauge transformations \cite{AGJL, AGJL2}, 
under Neveu-Schwarz-Neveu-Schwarz (NS-NS) transformations \cite{AIJ} and unifying it in a 
global picture in \cite{Adam}.

One of the remarkable results of these papers is that the NS-NS gauge transformations are much 
more complicated than the R-R or the massive gauge transformations, as they not only affect 
the supersymmetry background fields, but also the worldvolume fields living on the D-branes. 
The reason why this is so, is a mixture of Abelian and non-Abelian effects: already in the case of
a single (Abelian) D-brane, the Born-Infeld vector transforms with the pullback of the gauge
parameter $\Sigma_\mu$ of the NS-NS transformation of the Kalb-Ramond field $B_\mn$. However in 
the non-Abelian case this pullback is performed with $U(N)$ covariant derivatives, which under 
T-duality generates non-trivial transformation rules for the embedding scalars, in very much 
the same way as the dielectric couplings to higher-form R-R potentials appear in the action of 
multiple coinciding D-branes \cite{Myers}.   

This leads immediately to a second issue: where the R-R and massive transformations are merely a 
straightforward generalization of their Abelian counterparts in the single D-brane system, the 
non-Abelian NS-NS transformation rules involve non-trivial commutator terms that are not present 
in the Abelian case \cite{AGJL2, AIJ}.

Even though the explicit invariance of the non-Abelian Chern-Simons term of the effective 
worldvolume action under the NS-NS transformations is proven in \cite{AIJ}, 
the complicated transformation rules remain surprising and the aim 
of this letter is to study their structure in more detail. The main result of this paper is that 
the non-Abelian NS-NS transformations no longer have a $U(1)$ group structure, but that they 
intertwine non-trivially with the $U(N)$ Yang-Mills group of the worldvolume theory. We will show 
that part of the NS-NS transformations can be written as a $U(N)$ gauge transformation, but that 
there is also a non-trivial part that can not. We will construct explicitly the algebra
spanned by the NS-NS and the $U(N)$ transformations by computing the commutation relations and the 
Jacobi identities. 

The organization of this paper is as follows: in section 2 we will derive the NS-NS
transformation rules for the worldvolume fields, following the argument of \cite{AIJ}. In section 
3 we will construct and comment the NS-NS rules for matrix functions and composite objects such
as commutators and covariant derivative, needed in the next sections. In section 4 we will construct
the algebra formed by the NS-NS and $U(N)$ transformations and check the Jacobi identities in 
section 5. We also review some useful issues of non-commutative algebra in the appendix A. 
Finally, we summarize our results in the conclusions.  

\sect{The non-Abelian NS-NS transformation rules}

In this section we will review quickly the derivation of the NS-NS transformation rules for the
worldvolume fields, as done in \cite{AIJ}, based on the T-duality between D$p$- and 
D$(p-1)$-brane actions. The field content of a set of $N$ multiple coinciding D$q$-branes 
consists of a  $(q+1)$-dimensional Yang-Mills (Born-Infeld) vector $V_{a}$, that acts as 
the gauge field of the $U(N)$ symmetry group of the the system, and a set of $9-q$ matrix-valued 
transverse scalars $X^i$, that 
transform in the adjoint representation of $U(N)$. From the target space point of view, the latter 
have the interpretation of the non-Abelian embedding scalars of the D$p$-branes, while the former 
arises as the potential caused by the charged endpoints of open strings ending on the branes.

The T-duality between the D$p$-brane and the D$(p-1)$-brane system states that the physical content
of both theories is equivalent. This equivalence can be seen through an explicit mapping of the 
degrees of freedom of the two theories, that transforms one action in the other. We will denote 
the $9-p$ embedding scalars and the $(p+1)$-dimensional Born-Infeld vector of the D$p$-brane by
$Y^i$ and $\hat V_{\hat a}$, and the $10-p$ embedding scalars and the $p$-dimensional Born-Infeld 
vector of the D$(p-1)$-brane by $X^{\hat \imath}$ and $V_a$ respectively. Then the T-duality  
rules that relate the field contents of both theories after dualising in a worldvolume direction 
$x$ of the D$p$-brane are given by \cite{BR}
\bea
&& \hat V_a \longrightarrow V_a, 
\hsp{3cm}
Y^i \longrightarrow X^i, \nonumber \nnw
&& \hat V_x \longrightarrow X^\ux,
\hsp{2.85cm}
Y^{\ux} \ = \ \sigma^x,
\label{TdualWV}
\eea 
where we have decomposed $\hat V_{\hat a}$ and $X^{\hat \imath}$ as
$\hat V_{\hat a}= (\hat V_a, \hat V_x)$ and  $X^{\hat \imath}= (X^i, X^\ux)$ and $\sigma^x$ is the 
worldvolume coordinate of the D$p$-brane in the $x$-direction. The last equation is then merely
an expression of the fact that we write the actions in the static 
gauge, at least the direction in which the T-duality is performed, while the first two state 
that the BI vector components and the transverse scalars in directions different from the 
T-dualised one are the same in both actions. The non-trivial part of the T-duality rules is 
contained in the third equation, that matches the degrees of freedom of the two theories by 
mapping the $x$-component of the D$p$-brane BI vector with the extra embedding scalar of the 
D$(p-1)$-brane \cite{BR}.

Of course, consistency requires that not only the degrees of freedom of both actions match
according to (\ref{TdualWV}), but also their variations. In the Abelian limit this can easily 
be shown to be the case. We know that on the one hand the (Abelian) Born-Infeld vector  
$\hat V_{\hat a}$ transforms as a vector under general coordinate transformations 
$\hat \zeta^{\hat a}$ in the worldvolume (i.e. reparametrisations of the worldvolume directions), 
as a gauge field under $U(1)$ transformations $\hat \chi$ and with the pull-back of a shift
under NS-NS gauge transformations of the supergravity Kalb-Ramond field 
$\delta B_\hmn = 2\part_{[\hmu}\Sigma_{\hnu]}$:
\be
\delta \hV_\ha \ = \ \hzeta^\hb \, \partial_\hb \hV_\ha 
             \ + \ \partial_\ha \hzeta^\hb \, \hV_\hb
             \ - \ \partial_\ha \hchi
             \ - \ \Sigma_\hmu \,\partial_\ha Y^\hmu. 
\label{deltahV}
\ee
On the other hand, the Abelian embedding scalars transform as scalars under worldvolume coordinate 
transformations and as target space coordinates under target space diffeomorphisms $\delta x^\hmu
= - \xi^\hmu$:
\be
\delta X^\hmu \ = \ \zeta^b\,\partial_b X^\hmu  \  - \ \xi^\hmu. 
\ee   
It can easily be checked that indeed not only the field content transforms as in 
(\ref{TdualWV}), but also their variations dualise as
\bea
\delta \hat V_a \longrightarrow  \delta V_a,
\hsp{2cm}
\delta \hat V_x \longrightarrow \delta X^\ux, 
\hsp{2cm}
\delta Y^i \longrightarrow \delta X^i, 
\label{Tdualvar}
\eea 
provided that the transformation parameters map to each other  as ($\hmu = (\mu, \ux)$)
\bea
\begin{array}{lll}
\hzeta^a \longrightarrow \zeta^a,  \hsp{2cm}  
&  \Sigma_\mu \longrightarrow \Sigma_\mu, \hsp{1.8cm} 
& \hchi \longrightarrow \chi \ +\ \Sigma_\ux X^\ux,\\[.2cm]
\hzeta^x \longrightarrow \Sigma_\ux,
  & \Sigma_\ux \longrightarrow \xi^\ux.
\end{array}
\label{Tdualparameters}
\eea

The non-Abelian case is a bit more involved: here the non-Abelian Born-Infeld vector still 
transforms as a vector under worldvolume reparametrisations, but is now promoted to a $U(N)$ 
Yang-Mills gauge field and, more importantly, the pullback of the NS-NS parameter $\Sigma_\hmu$ 
needs now to be done through $U(N)$-covariant derivatives:
\be
\delta \hV_\ha \ = \ \hzeta^\hb\, \partial_\hb \hV_\ha 
             \ + \ \partial_\ha \hzeta^\hb \, \hV_\hb
             \ - \ \hD_\ha \hchi
             \ - \ \Sigma_\hmu \,\hD_\ha Y^\hmu. 
\label{deltahVNA}
\ee      
The transformation rules for the non-Abelian scalars $X^i$ get even more corrections: 
besides transforming as scalars and as coordinates under worldvolume and target space general  
coordinate transformations respectively, they also transform as adjoint scalars of $U(N)$ and, 
surprisingly, acquire also a transformation under the NS-NS transformation $\Sigma_\hmu$ 
\cite{AGJL2, AIJ}: 
\be
\delta X^\hmu \ = \ \zeta^b \partial_b X^\hmu \ - \ \xi^\hmu  
               \ + \ i [\chi, X^\hmu] \ + \ i \Sigma_\hrho [X^\hrho, X^\hmu]. 
\label{varXNA}
\ee
The last term is quite unexpected from the Abelian point of view, as it consists purely of a 
commutator. Yet it arises very naturally from the T-dualisation of the last term in the variation 
of $\delta V_x$ in (\ref{deltahVNA}), due to the covariant derivative in the pullback,
\be
\hD_x Y^i =  i [\hat V_ x, Y^i] \ \longrightarrow \ i [X^\ux, X^i],
\ee
as T-duality assumes $x$ to be an isometry direction and hence $\partial_x Y^i = 0$.
Note that this mechanism that generates these transformations is exactly the same as the one that 
generates the dielectric coupling terms in the Chern-Simons term of multiple coinciding D-branes 
\cite{Myers}. In a certain sense the NS-NS transformation rules can be though of as 
``dielectric gauge transformations''. It should  then be clear that 
the presence of this term is crucial for the transformation rules for the variations 
(\ref{Tdualvar}) to hold also in the non-Abelian case and hence for the consistency of the set-up.
It is precisely this term that will change the group structure of the non-Abelian NS-NS 
transformations.

\sect{More involved NS-NS transformation rules}

In the previous section we saw that form invariance under T-duality of the non-Abelian D-brane 
actions implies that a gauge transformation of the background Kalb-Ramond 
field\footnote{We omit hats on the indices 
henceforth, as we will not apply T-duality transformations in the rest of the paper. The notation 
should be self-explicatory} $\delta B_\mn = 2\part_{[\mu}\Sigma_{\nu]}$ in the target space
induces a simultaneous transformation of the embeddings 
scalars and the Born-Infeld vector in the worldvolume, according to
\bea
\delta B_\mn = 2 \partial_{[\mu}\Sigma_{\nu]}, 
\hsp{1.3cm}
\delta X^\mu = \ i \Sigma_\rho [X^\rho, X^\mu], 
\hsp{1.3cm}
\delta V_a = \ - \ \Sigma_\mu  D_a X^\mu, 
\label{Sigmatransf}
\eea
(for technical details on multiplication of algebra elements and non-Abelian functions, 
we refer to Appendix \ref{non-Abelianfunction}).

There are however some derived objects that do not transform as simple as the rules given above.
Although they will not yield new insights in the group structure of the NS-NS algebra, that fact
that they obey the same algebra will make a strong case for the consistency of our approach.

In general, a non-Abelian function $\Phi(X)$ of the embedding coordinates $X^\mu$ 
transforms under NS-NS gauge transformations as
\begin{equation}
  \delta \Phi (X) \ = \ \partial_{\mu} \Phi (X)  \delta X^\mu
                 \ =  \ \partial_{\mu} \Phi (X) \ i \Sigma_\rho [X^\rho, X^\mu]
                \ =\  i \; \Sigma_\mu  [ X^\mu , \Phi (X) ],
\end{equation}
where in the last step we used the matrix identity (\ref{[Phi,X]}).
 
The variation of the $U(N)$ covariant derivative  $D_a X^\mu$ is given by 
\bea
\delta D_{a} X^\mu 
  &=& \partial_a(\delta X^\mu) \ + \ i[\delta V_a, X^\mu] \ + \ i [V_a, \delta X^\mu] \nnw
  &=& \partial_a(i\Sigma_\rho [X^\rho, X^\mu]) 
       \ + \ i[-\Sigma_\rho D_a X^\rho, X^\mu]
       \ + \ i [V_a, i\Sigma_\rho [X^\rho, X^\mu]] \nnw
  &=& \partial_a(i\Sigma_\rho [X^\rho, X^\mu]) 
      \ +\ i\Sigma_\rho [\partial_a X^\rho, X^\mu]
    \ + \ i\Sigma_\rho[ X^\rho, \partial_a X^\mu] 
      \ + \ i\Sigma_\rho  [ D_a X^\rho, X^\mu]  \nnw
 &&  \ + \ i[\Sigma_\rho, X^\mu] D_a X^\rho
       \ - \ \Sigma_\rho [V_a, [X^\rho, X^\mu]]
       \ - \ [V_a, \Sigma_\rho][X^\rho, X^\mu] \nnw
  &=& i \Sigma_\rho  [X^\rho, D_{a} X^\mu] \ + \
      i[X^\mu, \Sigma_\lambda] D_a X^\lambda -i [X^\mu, X^\lambda] D_{a} \Sigma_\lambda,
\eea
where the last two terms can be unified as
\bea
\delta D_{a} X^\mu  =  i \Sigma_\rho  [X^\rho, D_{a} X^\mu] \ + \
      2i[X^\mu, X^\lambda]  \partial_{[\lambda}\Sigma_{\nu]}  D_{a} X^\nu.
\label{varDX}
\eea

The variation of the commutator $[X^\mu, X^\nu] $ can be calculated as
\bea
\delta [X^\mu, X^\nu] 
   &=& [\delta X^\mu, X^\nu] \ + \ [X^\mu, \delta X^\nu]  \\[.2cm]
   &=& [i\Sigma_\rho [X^\rho, X^\mu], X^\nu]  
          \ + \  [X^\mu, i\Sigma_\rho [X^\rho, X^\nu]]  \nnw
   &=& i\Sigma_\rho [[X^\rho, X^\mu], X^\nu]  
                  \ + \ i[\Sigma_\rho, X^\nu] [X^\rho, X^\mu] \nnw
   && \hsp{1cm} 
        + \  [X^\mu, i\Sigma_\rho] [X^\rho, X^\nu] 
       \ + \ i\Sigma_\rho [X^\mu, [X^\rho, X^\nu]]\nnw
   &=& i\Sigma_\rho[X^\rho, [X^\mu, X^\nu]] \ + \ i[X^\mu, \Sigma_\rho][X^\rho, X^\nu]
        \ - \ i [X^\mu, X^\rho][\Sigma_\rho, X^\nu]. \nonumber
\eea
Again the last two terms can be taken together as
\be
\delta [X^\mu, X^\nu] \ = \ i\Sigma_\rho[X^\rho, [X^\mu, X^\nu]] 
         \ + \ 2i  [X^\mu, X^\rho] \partial_{[\rho}\Sigma_{\lambda]}[X^\lambda, X^\nu].
\label{var[XX]}
\ee
Finally, it will be useful to compute also the variation of a double commutator 
$[[X^\mu, X^\nu], X^\lambda]$. Using the above result we have that
\bea
\delta [[X^\mu, X^\nu], X^\lambda] 
   &=& i \Sigma_\rho\Bigl[X^\rho, [[X^\mu, X^\nu], X^\lambda]\Bigr] 
   \ - \ i \Bigl[ [X^\mu, X^\nu], X^\rho\Bigr][\Sigma_\rho, X^\lambda]
 \\ [.3cm]
  && \hsp{-3cm}
+ \ i \Bigl[ [X^\mu, X^\nu], \Sigma_\rho \Bigr][X^\rho, X^\lambda]
     \ - \ i \Bigl[[  [X^\mu, X^\rho][\Sigma_\rho, X^\nu], X^\lambda \Bigr]
      \ + \ i \Bigl[[  [X^\mu, \Sigma_\rho ][X^\rho, X^\nu], X^\lambda \Bigr].
\nonumber
\eea

We therefore see that the NS-NS transformations treat embedding scalars, their covariant 
derivatives and their commutators on different footing, even if they all sit in the adjoint of 
$U(N)$. A priori there is of course no reason to expect that the NS-NS gauge symmetry would respect
the same structures as the $U(N)$ group. In the next section however we will show that there is a 
certain relation between the two groups.

\sect{The algebra of NS-NS transformations }

A first hint of the group structure of the NS-NS transformations (\ref{Sigmatransf})
comes from realising that there is a class of transformations that leave $B_\mn$ invariant, but 
act non-trivially on the worldvolume fields. Indeed, taking the NS-NS parameter to be exact,
$\Sigma_\mu = \partial_\mu \Lambda$, the transformation rules (\ref{Sigmatransf}) take 
the form
\bea
 \delta B_\mn = 0, 
\hsp{1.3cm}
\delta X^\mu = \ i \partial_\rho \Lambda  [X^\rho, X^\mu], 
\hsp{1.3cm}
\delta V_a = \ - \ \partial_\mu \Lambda  D_a X^\mu. 
\eea

The only non-trivial symmetry in the worldvolume theory is the $U(N)$ gauge group, such that one
would expect the above transformations to be $U(N)$ gauge symmetries. Indeed, with the aid of 
(\ref{[Phi,X]}) and (\ref{D=d}), these rules can be rewritten as
\bea
\delta B_\mn = 0, 
\hsp{1.3cm}
\delta X^\mu = \ i [ \Lambda, X^\mu], 
\hsp{1.3cm}
\delta V_a = \ - \ D_a \Lambda,
\eea 
i.e. as $U(N)$ gauge transformation with parameter $\Lambda$. Also the more complicated 
NS-NS transformation rules for the commutator (\ref{var[XX]}) and  the covariant derivative 
(\ref{varDX}) reduce to the standard transformation rules for objects in the adjoint representation
of $U(N)$:
\bea
\delta [X^\mu, X^\nu] \ = \ i [\Lambda,  [X^\mu, X^\nu]], 
\hsp{2cm}  
\delta D_a X^\mu \ = \ i [\Lambda,  D_a X^\mu]. 
\eea  

We therefore see that the pure gauge part of the NS-NS transformations is in fact a $U(N)$ symmetry.
However a general NS-NS transformation can not be written as a part of the $U(N)$-algebra and it 
would be interesting to analyse the complete structure of the intertwining NS-NS and $U(N)$ 
transformations. We will therefore calculate the commutators of the NS-NS variations amongst each 
other, NS-NS with $U(N)$ and $U(N)$ with itself.

The last case is in fact trivial, as we are considering the commutation rules of the $U(N)$ 
(sub-)algebra,
\be
[\delta_{\chi_1}, \delta_{\chi_2}] = \delta_{\chi_3} \hsp{2cm}
{\rm with} \hsp{1cm} \chi_3 = i[\chi_1, \chi_2].
\ee
Less trivial are the commutators involving NS-NS transformations. The mixed $U(N)$ and NS-NS
commutator, acting on the scalars, is given by
\bea
[\delta_\Sigma, \delta_\chi] X^\mu 
 &=& \delta_\Sigma\Bigl(i[\chi, X^\mu]\Bigr)
      \ -\ \delta_\chi \Bigl(i\Sigma_\rho[X^\rho, X^\mu]\Bigr) \\[.2cm]
 &=& -\Sigma_\rho[X^\rho, [\chi, X^\mu]] + [\chi, X^\rho][\Sigma_\rho, X^\mu]
         - [\chi, \Sigma_\rho][X^\rho , X^\mu] + [\chi, \Sigma_\rho [X^\rho, X^\mu]].
\nonumber
\eea
Using the ($U(N)$) Jacobi identities and the decomposition rules for nested commutators, we see
that the above commutator can be written as a $U(N)$ transformation
\be
[\delta_\Sigma, \delta_\chi] X^\mu \ = \ i[\tilde \chi, X^\mu]\hsp{2cm}
{\rm with} \hsp{1cm} \tilde \chi = i\Sigma_\rho[X^\rho, \chi].
\ee 
The same commutator can be checked acting on the Born-Infeld vector $V_a$ to yield
\be
[\delta_\Sigma, \delta_\chi] V_a \ = \ - \ D_a{\tilde \chi},
\ee
with $\tilde \chi$ given by the same expression. 

Finally, the commutator of two NS-NS transformations is non-trivial as well. Acting on the scalars
we find that
\bea
&&  \hsp{-1cm}
[\delta_{\Sigma^{(1)}},\delta_{\Sigma^{(2)}}] X^\mu\ =\
  - \Sigma^{(1)}_\lambda  [X^\lambda, \Sigma^{(2)}_\rho][X^\rho, X^\mu] 
   \ + \ \Sigma^{(2)}_\lambda  [X^\lambda, \Sigma^{(1)}_\rho][X^\rho, X^\mu] \\ [.2cm]
&& + \ i \Sigma^{(2)}_\rho \left(i \Sigma^{(1)}_\lambda [X^\lambda, [X^\rho, X^\mu]] 
     \ + \ i [X^\rho, \Sigma^{(1)}_\lambda][X^\lambda, X^\mu]
     \ - \ i [X^\rho, X^\lambda ][\Sigma^{(1)}_\lambda, X^\mu] \right) \nnw
&& + \ i \Sigma^{(1)}_\rho \left(i \Sigma^{(2)}_\lambda [X^\lambda, [X^\rho, X^\mu]] 
     \ + \ i [X^\rho, \Sigma^{(2)}_\lambda][X^\lambda, X^\mu]
     \ -  \ i [X^\rho, X^\lambda ][\Sigma^{(2)}_\lambda, X^\mu] \right). \nonumber
\eea
Given that the first and the second term cancel the seventh and the fifth respectively,
this expression can be simplified to 
\bea 
[\delta_{\Sigma^{(1)}},\delta_{\Sigma^{(2)}}] X^\mu
&=& \Sigma^{(1)}_\lambda  \Sigma^{(2)}_\rho 
        [X^\mu, [X^\lambda, X^\rho ]] 
\ + \  [X^\mu, \Sigma^{(1)}_\lambda  \Sigma^{(2)}_\rho ] [X^\lambda, X^\rho ] \nnw
&=& i \Bigl[i\Sigma^{(1)}_\lambda  \Sigma^{(2)}_\rho  [X^\lambda, X^\rho ], X^\mu \Bigr]. 
\eea
In other words, the commutator of two NS-NS variations is again a $U(N)$ transformation with 
parameter 
\be
\bar \chi = i\Sigma^{(1)}_\lambda  \Sigma^{(2)}_\rho   [X^\lambda, X^\rho ].
\ee
Again the same result is obtained for the commutator acting on $V_a$:
\bea 
[\delta_{\Sigma^{(1)}},\delta_{\Sigma^{(2)}}] V_a
      = \ - \ D_a \left(i\Sigma^{(1)}_\lambda  \Sigma^{(2)}_\rho   [X^\lambda, X^\rho] \right).
\eea

We find therefore that the $U(N)$ and the NS-NS transformations form  a larger algebra given by
\bea
\begin{array}{ll}
[\delta_{\chi_1}, \delta_{\chi_2}] = \delta_{\chi_3} \hsp{1cm} & 
\hsp{1cm} {\rm with} \hsp{.4cm} \chi_3 =  i[\chi_1, \chi_2],
\\ [.3cm]
[\delta_{\Sigma}, \delta_{\chi}] = \delta_{\tilde \chi}  &
\hsp{1cm} {\rm with} \hsp{.4cm} 
    \tilde \chi = i\Sigma_\rho[X^\rho, \chi],
\\ [.3cm]
[\delta_{\Sigma_1}, \delta_{\Sigma_2}] =  \delta_{\bar \chi} &  
\hsp{1cm} {\rm with} \hsp{.4cm} 
     \bar\chi = i\Sigma^{(1)}_\lambda \Sigma^{(2)}_\rho [X^\lambda, X^\rho].
\end{array}
\label{algebra}
\eea

It is worthwhile to check the above algebra on the more complicated transformation rules for 
the commutator and the covariant derivative. As we mentioned, it will not yield many new insights,
but it will give us more confidence in the derived results. 

The calculation simplifies greatly if we write in general the NS-NS transformation rules as
\be
\delta Z \ = \ \delta^{0} Z \ + \ \delta^*Z,
\ee
where $\delta^0 Z$ is the ``standard part''
\be
\delta^0 Z = i \Sigma_\rho [X^\rho, Z],
\ee 
for any $Z$, and $\delta^*Z$ is the correction terms that appear for $Z$ being $[X^\mu, X^\nu]$ 
or $D_aX^\mu$,
\bea
&& \delta^* [X^\mu, X^\nu] \ = \ i[X^\mu, \Sigma_\rho][X^\rho, X^\nu]
        \ - \ i [X^\mu, X^\rho][\Sigma_\rho, X^\nu], \nnw
&& \delta^* D_a X^\mu \ =\ i[X^\mu, \Sigma_\lambda]  D_a X^\lambda 
                        - i [X^\mu, X^\lambda] D_{a} \Sigma_\lambda.
\eea
The trick now is to realise that the $\delta^*$ part commutes with both $\delta^0$ and $U(N)$ 
variations. Indeed, we have that
\bea
\delta^*(\delta_\chi Z) \ =\ \delta^* \Bigl( i[\chi, Z]\Bigr)
                        \ =\ i [\delta^*\chi, Z] \ + \ i [\chi, \delta^*Z]
                        \ = \ \delta_\chi(\delta^* Z),
\eea 
as the correction terms  $\delta^*$ vanish for $\chi$. Hence the mixed commutator 
$[\delta_\Sigma, \delta_\chi]$ reduces to the standard part
\be
[\delta_\Sigma, \delta_\chi] Z \ = \ [\delta^0, \delta_\chi]Z \ + \ [\delta^*, \delta_\chi]Z 
                               \ = \ \delta_{\tilde \chi} Z,
\ee 
as the second term vanishes identically. Similarly the commutator 
$[\delta_{\Sigma_1}, \delta_{\Sigma_2}]$ simplified to
\be
[\delta_{\Sigma_1}, \delta_{\Sigma_2}] Z 
\ = \  [\delta^0, \delta^0]Z \ + \ [\delta^*, \delta^*]Z.
\ee
The first term in the right-hand side is again the standard commutator $\delta_{\bar \chi}Z$, and
although the second term has to be calculated for each case separately, it can be shown to
vanish identically for both $[X^\mu, X^\nu]$ and $D_aX^\mu$. Hence we see indeed that the algebra 
(\ref{algebra}) is also satisfied for these objects, in spite of their involved transformation 
rules.

\sect{Jacobi identities}

Finally, in order to ensure the consistency of the algebra (\ref{algebra}), we will check the 
Jacobi identities. There are four identities to be checked, and from the structure of the
algebra it is clear that all will result in a $U(N)$ transformation. We will prove that the Jacobi 
identities are satisfied by showing that the resulting $U(N)$ transformations have zero parameter.

Indeed, the different identities, acting on a generic object $Z$, yield  $U(N)$ 
transformations
\bea
&& [\delta_{\chi_1}, [\delta_{\chi_2},\delta_{\chi_3}]] Z 
   \ + \ [\delta_{\chi_2}, [\delta_{\chi_3},\delta_{\chi_1}]]Z 
   \ + \ [\delta_{\chi_3}, [\delta_{\chi_1},\delta_{\chi_2}]] Z
   \ = \ i [\chi_0, Z], \nnw
&& [\delta_{\Sigma}, [\delta_{\chi_1},\delta_{\chi_2}]]Z
   \ + \ [\delta_{\chi_2}, [\delta_\Sigma,\delta_{\chi_1}]] Z
   \ + \ [\delta_{\chi_1}, [\delta_{\chi_2},\delta_{\Sigma}]] Z
   \ = \ i [\tilde \chi, Z], \nnw
&& [\delta_{\Sigma_1}, [\delta_{\Sigma_2},\delta_{\chi}]]Z
   \ + \ [\delta_{\chi}, [\delta_{\Sigma_1},\delta_{\Sigma_2}]] Z
   \ + \ [\delta_{\Sigma_2}, [\delta_{\chi},\delta_{\Sigma_1}]] Z
   \ = \ i [\bar \chi, Z], \nnw
&& [\delta_{\Sigma_1}, [\delta_{\Sigma_2},\delta_{\Sigma_3}]]Z
   \ + \ [\delta_{\Sigma_3}, [\delta_{\Sigma_1},\delta_{\Sigma_2}]] Z
   \ + \ [\delta_{\Sigma_2}, [\delta_{\Sigma_3},\delta_{\Sigma_1}]] Z
   \ = \ i [\hat \chi, Z],
\eea
with the parameters $\chi_0$, $\tilde \chi$,  $\bar \chi$ and  $\hat \chi$ given by  
\bea
&& \chi_0 \ = \ [\chi_1, [\chi_2,\chi_3]] 
   \ + \ [{\chi_2}, [{\chi_3},{\chi_1}]]
   \ + \ [{\chi_3}, [{\chi_1},{\chi_2}]] \ = \ 0, \nnw
&& \tilde \chi \ = \ \Sigma_\rho [[\chi_1, \chi_2], X^\rho] 
               \ + \ [\chi_1, X^\rho][\Sigma_\rho, \chi_2] 
               \ -\ [\chi_1, \Sigma_\rho][X^\rho , \chi_2] \nnw
&& \hsp{3,8cm} 
                 - \ [\chi_2, \Sigma_\rho[X^\rho, \chi_1]] 
               \ + \ [\chi_1, \Sigma_\rho[ X^\rho, \chi_2]],\nnw 
&& \bar \chi \ = \ - \Sigma^{(1)}_\lambda [X^\lambda, \Sigma^{(2)}_\rho[X^\rho, \chi]]
              \ + \ \Sigma^{(2)}_\lambda [X^\lambda, \Sigma^{(1)}_\rho[X^\rho, \chi]] 
             \  - \ [\chi, \Sigma^{(1)}_\lambda \Sigma^{(2)}_\rho[X^\lambda, X^\rho]] \nnw
&&  \hsp{3,8cm} 
               + \ \Sigma^{(2)}_\rho [X^\rho, X^\lambda][\Sigma^{(1)}_\lambda , \chi] 
             \ - \ \Sigma^{(2)}_\rho [X^\rho, \Sigma^{(1)}_\lambda ][ X^\lambda , \chi]\nnw
&& \hsp{3,8cm} - \ \Sigma^{(1)}_\rho [X^\rho, X^\lambda][\Sigma^{(2)}_\lambda , \chi] 
             \ + \ \Sigma^{(1)}_\rho [X^\rho, \Sigma^{(2)}_\lambda ][ X^\lambda , \chi],\nnw
&& \hat \chi
   \ = \ -\Sigma^{(1)}_\nu [X^\nu,\Sigma^{(2)}_\lambda\Sigma^{(3)}_\rho[X^\lambda, X^\rho]]
  \ + \ \Sigma^{(2)}_\lambda\Sigma^{(3)}_\rho[X^\lambda, X^\nu][\Sigma_\nu^{(1)}, X^\rho] \nnw
&& \hsp{3,8cm}
  \ - \ \Sigma^{(2)}_\lambda\Sigma^{(3)}_\rho[X^\lambda,\Sigma_\nu^{(1)} ][X^\nu, X^\rho] \nnw
&& \hsp{0,8cm}
  \ - \ \Sigma^{(3)}_\rho [X^\rho,\Sigma^{(1)}_\nu\Sigma^{(2)}_\lambda[X^\nu, X^\lambda]]
  \ + \ \Sigma^{(1)}_\nu\Sigma^{(2)}_\lambda[X^\nu, X^\rho][\Sigma_\rho^{(3)}, X^\lambda] \nnw
&& \hsp{3,8cm}
  \ - \ \Sigma^{(1)}_\nu\Sigma^{(2)}_\lambda[X^\nu,\Sigma_\rho^{(3)} ][X^\rho, X^\lambda] \nnw
&& \hsp{0,8cm}
  \ - \ \Sigma^{(2)}_\lambda [X^\lambda,\Sigma^{(3)}_\rho\Sigma^{(1)}_\nu[X^\rho, X^\nu]]
  \ + \ \Sigma^{(3)}_\rho\Sigma^{(1)}_\nu[X^\rho, X^\lambda][\Sigma_\lambda^{(2)}, X^\nu] \nnw
&& \hsp{3,8cm}
  \ - \ \Sigma^{(3)}_\rho\Sigma^{(1)}_\nu[X^\rho,\Sigma_\lambda^{(2)} ][X^\lambda, X^\nu].
\eea
The parameter $\chi_0$ vanishes identically due to the Jacobi identity of $U(N)$, but 
with a little matrix algebra also the other expressions can quite easily be shown 
to be proportional to the $U(N)$ Jacobi identities and therefore to vanish identically:
\bea
&& \tilde \chi \ = \ \Sigma_\rho \Bigl(
                [[\chi_1, \chi_2], X^\rho] 
                 +  [[ X^\rho, \chi_1], \chi_2]
                 +  [[ \chi_2, X^\rho], \chi_ ] \Bigr) \ = \ 0, \nnw 
&& \bar \chi \ = \ - \Sigma^{(1)}_\lambda \Sigma^{(2)}_\rho 
            \Bigl( [X^\lambda, [X^\rho, \chi]] + [ \chi, [ X^\lambda, X^\rho]]
                           +   [X^\rho, [ \chi,  X^\lambda]] \Bigr) \ = \ 0,  \nnw
&& \hat \chi
   \ = \ -\Sigma^{(1)}_\nu \Sigma^{(2)}_\lambda\Sigma^{(3)}_\rho 
           \Bigl( [X^\nu,[X^\lambda, X^\rho]] + [X^\rho, [X^\nu,X^\lambda]]
                    + [X^\lambda, [X^\rho, X^\nu]] \Bigr) \ = \ 0.
\eea

\sect{Conclusions}

{}From  the derivation performed in \cite{AGJL2, AIJ} it is known that the background gauge 
transformation of the Kalb-Ramond field $\delta B_\mn = 2\partial_{[\mu}\Sigma_{\nu]}$ induces
non-trivial, ``dielectric'', transformations of the worldvolume field content of the non-Abelian 
action of a system of multiple coinciding D-branes. These NS-NS transformations acting on the 
embedding scalars take the form of a pure commutator, $\delta X^\mu = i\Sigma_\rho [X^\rho, X^\mu]$,
and arises from T-dualising the transformation rules for the Born-Infeld vector $\delta V_a =
-\Sigma_\mu D_aX^\mu$, in the same way as the dielectric couplings arise in the Myers action.    

It has been shown already in \cite{AIJ} that the Chern-Simons action of the system of multiple 
D-branes is invariant under the NS-NS transformations of both the background and the worldvolume 
fields. In this letter we have shown that the group structure of the NS-NS transformations is more 
involved than in the Abelian case, due to the fact that the non-Abelian NS-NS transformations 
rules imply a non-trivial mixture with the $U(N)$ Yang-Mills symmetry. 

A first hint of this can be seen if we take the NS-NS parameter to be exact, 
$\Sigma_\mu = \partial_\mu \Lambda$. In that case, the supergravity part is untouched, 
$\delta B_\mn = 0$, but the worldvolume fields do transform non-trivially under a $U(N)$ 
gauge transformation with parameter $\Lambda$. In other words, the exact part of the NS-NS 
transformations can be written as a part of the $U(N)$ Yang-Mills symmetry of the worldvolume
theory. 

However, this reduction to $U(N)$ variations can not be done for a general NS-NS transformation
with arbitrary parameter $\Sigma_\mu$ and the full group structure of the intertwining $U(N)$ and 
NS-NS transformations is revealed by the full algebra  
\bea
\begin{array}{ll}
[\delta_{\chi_1}, \delta_{\chi_2}] = \delta_{\chi_3} \hsp{1cm} & 
\hsp{1cm} {\rm with} \hsp{.6cm} \chi_3 =  i[\chi_1, \chi_2],
\\ [.3cm]
[\delta_{\Sigma}, \delta_{\chi}] = \delta_{\tilde \chi}  &
\hsp{1cm} {\rm with} \hsp{.6cm} 
    \tilde \chi = i\Sigma_\rho[X^\rho, \chi],
\\ [.3cm]
[\delta_{\Sigma_1}, \delta_{\Sigma_2}] =  \delta_{\bar \chi} &  
\hsp{1cm} {\rm with} \hsp{.6cm} 
     \bar\chi = i\Sigma^{(1)}_\lambda \Sigma^{(2)}_\rho [X^\lambda, X^\rho].
\end{array}
\label{algebra2}
\eea
We see that the $U(N)$ algebra of the Yang-Mills is a non-trivial sub-algebra of the full algebra, 
which also involves non-trivial commutators between NS-NS and $U(N)$ transformations and between
NS-NS amongst each other. The surprising issue however is that all the resulting commutators turn 
out to be $U(N)$ transformations, the inherent gauge symmetry of the worldvolume theory. In a 
certain sense, the $U(N)$ symmetry ``non-Abelianises'' the other gauge symmetries present (in this
case the NS-NS transformations), but only in a very mild way: modulo a $U(N)$ transformation,
they still all behave as if they were $U(1)$ gauge symmetries in the Abelian theory.  

A few comments are in order: first, as already mentioned in \cite{AIJ}, there seems to be a 
remarkable difference between the NS-NS gauge transformations of $B_\mn$ and the R-R 
transformations of $C_\mn$, 
at least at the level of the worldvolume theory, in spite of the fact that $B_\mn$ and $C_\mn$ 
form a doublet under S-duality. The reason for this is of course that S-duality takes us out of 
the perturbative regime in which we can trust the description we have been working with. However
it should be clear that if we believe in S-duality in Type IIB string theory, there should be 
a ``non-Abelianisation'' (in the sense of (\ref{algebra2})) of the R-R gauge transformations, not 
only of the R-R two-form $C_\mn$, but via T-duality of all other R-R potentials as well.

Secondly, it is well known that (parts of the) NS-NS transformations mix with (part of the) general 
coordinate transformations under T-duality. Knowing on the one hand that a satisfactory description 
of general coordinate transformations in non-commutative (or matrix-valued) geometry is still an 
open issue, and keeping in mind  the non-trivial structure of the non-Abelian NS-NS transformations
on the other hand, it will be clear that one might find interesting and 
surprising results on non-commutative geometry and gauge transformations if one tried to T-dualise 
the algebra (\ref{algebra2}). We leave these ideas for future investigations.

\appendix 
\renewcommand{\theequation}{\Alph{section}.\arabic{equation}}
\sect{The non-Abelian formalism}
\label{non-Abelianfunction}

There are many ways to generalise functions $\Phi(x)$ of the Abelian coordinates $x^\mu$ to 
matrix functions $\Phi(X)$ of the matrix-valued coordinates $X^\mu$. One of the most common 
prescriptions (and the one traditionally used in the context of multiple coinciding D-brane 
systems) 
is the symmetrised prescription, through the non-Abelian Taylor expansion of $\Phi(X)$,
\be
\Phi(X) =  \sum_{n=0}^{\infty} \frac{1}{n!} \left. 
     \partial_{\mu_1} \ldots  \partial_{\mu_n} \Phi \right|_{x=0} 
                X^{\mu_1}  \ldots  X^{\mu_n},
\label{eqn:taylor_expansion}
\ee
where the matrix multiplication is taken to be the symmetrised product of $n$ Lie algebra 
elements $A_1, \ldots, A_n$, 
\be
A_1 \ldots  A_n= \frac{1}{n!} \sum_{\sigma \in S_n} A_{\sigma(1)} \ldots A_{\sigma(n)}. 
\ee
As the product between $X$'s is not defined within the structure of the algebra, strictly speaking
$\Phi (X)$ is not an element of the $u(N)$ algebra. However with the above definitions, $\Phi (X)$ 
becomes an element of the tensor algebra $T(u(N))$ of $u(N)$. In addition to this, the commutator
structure can be transferred to  $T(u(N))$, imposing the following equivalence relation
\begin{equation}
A B - B A \sim [ A , B ],
\label{eqn:equivalence_relation}
\end{equation}
yielding the so-called the universal enveloping algebra $U(u(N))$ of $u(N))$. Note that the 
symmetrised prescription has to be imposed on $U(u(N))$, rather then on $T(u(N))$, due to the 
equivalence relation (\ref{eqn:equivalence_relation}), as otherwise all commutators of the Lie 
algebra would vanish.

The non-Abelian functions defined in this way have a series of useful properties. For instance, 
the variation of the non-Abelian functions is given by
\be
\delta\Phi (X) 
   \ = \ \sum_{n=0}^{\infty} \sum_{i=1}^{n} \left. 
           \partial_{\mu_1} \ldots  \partial_{\mu_n} \Phi \right|_{x=0} 
              X^{\mu_1}  \ldots  \delta X^{\mu_i}  \ldots  X^{\mu_n} 
  \ = \ \partial_{\mu} \Phi (X)  \delta X^\mu,
\ee
where we used $\partial_{\mu} \Phi (X)$ as a shorthand for its non-Abelian Taylor expansion.
Given that the scalars transform under $u(N)$ gauge transformations as 
$\delta X^\mu = i[\chi, X^\mu]$, the variation of the non-Abelian function $\Phi (X)$ can then be 
written as
\be
\delta_{\chi} \Phi (X) \ = \ \partial_{\mu} \Phi (X)  i [\chi, X^\mu]
                       \ = \ i[\chi, \Phi (X)],
\ee
where in the last step we used the properties of the symmetrised prescription and the commutator
to prove that
\be
[\Phi(X), X^\mu] = \partial_\rho \Phi (X) [X^\rho, X^\mu].
\label{[Phi,X]}
\ee
In the same way we can also define a covariant derivative of $\Phi (X)$ via
\begin{equation}
D_{a} \Phi (X) \ = \ \partial_{a} \Phi (X) + i \; [ V_{a} , \Phi(X) ]
              \ = \ \partial_{\mu} \Phi (X)  D_{a} X^\mu.
\label{D=d}
\end{equation}
Similarly the commutator of two non-Abelian functions is given by
\begin{equation}
[ \Phi_1 (X) , \Phi_2 (X) ] 
       = i [X^\mu, X^\nu]  \partial_{\mu} \Phi_1 (X)  \partial_{\nu} \Phi_2 (X).
\label{eqn:commutator_Phis}
\end{equation}
It can easily be checked that this definition satisfied the Jacobi identity
\begin{equation}
[ \Phi_1, [ \Phi_2, \Phi_3 ] ] 
      + [ \Phi_2, [ \Phi_3, \Phi_1 ] ] 
          + [ \Phi_3, [ \Phi_1, \Phi_2 ] ] = 0.
\ee

\noindent
{\bf Acknowledgements}\\
The authors wish to thank J. Adam and J. Mas for useful discussions.
The work of B.J. is partially supported by the Spanish Ministerio de Ciencia e Innovaci\'on 
under contract FIS 2007-63364, by the Junta de Andaluc\'{\i}a group FQM 101 and the Proyecto de
Excelencia P07-FQM-03048.



\begin{thebibliography}{99}

\bibitem{Witten} E. Witten, Nucl. Phys. B460 (1996) 335, hep-th/9510135.

\bibitem{Tseytlin2} A.A. Tseytlin, {\it Born-Infeld action, supersymmetry and
string theory}, hep-th/9908105.

\bibitem{Douglas} M. Douglas, {\it Branes within Branes}, hep-th/9512077.

\bibitem{GHT} M. Green, C. Hull, P. Townsend, Phys. Lett. B382 (1996) 65, hep-th/9604119.

\bibitem{Dorn} H. Dorn, Nucl. Phys. B494 (1997) 105, hep-th/9612120.

\bibitem{Tseytlin} A. Tseytlin, Nucl. Phys. B501 (1997) 41, hep-th/9701125.

\bibitem{Douglas2} M. Douglas, Adv. Theor. Math. Phys. 1 (1998) 198, hep-th/9703056.

\bibitem{Hull} C. Hull, JHEP 9810 (1998) 011, hep-th/9711179.

\bibitem{GM} M. Garousi, R. Myers, Nucl. Phys. B542 (1999) 73, hep-th/9809100.

\bibitem{TvR} W. Taylor, M. Van Raamsdonk, Nucl. Phys. B573 (2000) 703, hep-th/9910052.

\bibitem{Myers} R. Myers, JHEP 9912 (1999) 022, hep-th/9910053.

\bibitem{Ciocarlie} C. Ciocarlie, JHEP 0107 (2001) 028, hep-th/0105253.

\bibitem{AGJL} J. Adam, J. Gheerardyn, B. Janssen, Y. Lozano, Phys. Lett. B589 (2004) 59, 
              hep-th/0312264.

\bibitem{AGJL2} J. Adam, J. Gheerardyn, B. Janssen, Y. Lozano, {\it On the gauge invariance of the 
non-Abelian Chern-Simons action for D-branes}, in N. Alonso et al (Ed.), {\sl Beyond General 
Relativity},  UAM Ediciones 119 (2007), 87-90, hep-th/0501206.


\bibitem{AIJ} J. Adam, I.A. Ill\'an, B. Janssen, JHEP 0510 (2005) 022, hep-th/0507198.

\bibitem{Adam} J. Adam, JHEP 0604 (2006) 007, hep-th/0511191. 

\bibitem{BR} E. Bergshoeff, M. de Roo, Phys. Lett. B380 (1996) 265, hep-th/9603123.


\end{thebibliography}
\end{document}